\documentclass[a4paper,11pt]{article}

\usepackage[utf8]{inputenc}
\usepackage[T1]{fontenc}
\usepackage[english]{babel}

\usepackage{indentfirst}
\usepackage{amsmath}
\usepackage{amssymb}
\usepackage{graphicx}
\usepackage{subfigure}
\usepackage{psfrag}{}
\usepackage{color}
\usepackage{pgf}

\allowhyphens

\newcommand{\eps}{\varepsilon}

\newcommand{\ordo}{\mathcal{O}}

\newcommand{\Pe}{\mathcal{P}}

\newcommand{\ket}[1]{{\left|#1\right\rangle}}
\newcommand{\bra}[1]{{\left\langle #1\right|}}

\newcommand{\skalarszorzat}[2]{{\langle #1 | #2 \rangle}}

\newcommand{\vev}[1]{\left\langle #1 \right\rangle}

\usepackage{ifpdf}

\ifpdf
\usepackage{epstopdf}
\usepackage[pdftex,ps2pdf,dvips,colorlinks,urlcolor=blue,citecolor=blue,linkcolor=blue]{hyperref}
\else
\usepackage[hypertex,ps2pdf,dvips,colorlinks,urlcolor=blue,citecolor=blue,linkcolor=blue]{hyperref}
\fi

\makeatother

\begin{document}

\numberwithin{equation}{section}

\author{Bal\'azs Pozsgay\\
Institute for Theoretical Physics, Universiteit van Amsterdam\\
Science Park 904, Postbus 94485, 1090 GL Amsterdam, The Netherlands
}
\title{Local correlations in the 1D Bose gas\\ from a scaling limit of
  the XXZ chain}
\date{August 2011}

\maketitle

\abstract{We consider the $K$-body local correlations in the
  (repulsive) 1D Bose
  gas for general $K$, both at finite size and in the thermodynamic limit. Concerning
  the latter  we develop a multiple integral formula which applies for
  arbitrary states of the system with a smooth distribution of Bethe
  roots, including the ground state and finite 
  temperature Gibbs-states. In the cases $K\le 4$ we perform the
  explicit factorization of the multiple integral. In the case of
  $K=3$ we obtain the recent result of Kormos et.al., whereas our
  formula for $K=4$ is new. Numerical results are presented as well.
}


\section{Introduction}

The delta-function interacting 1D Bose gas (also known as the
Lieb-Liniger model or the Quantum Nonlinear Schr\"odinger equation) is
one of the oldest and most important integrable models. Its study goes
back to the papers \cite{Lieb-Liniger,Lieb2} where it was shown that
the spectrum can be obtained by the Bethe Ansatz \cite{XXX}. The
thermodynamical properties of the model were determined in
\cite{YangYang2} using the method nowadays known as the Thermodynamical Bethe
Ansatz (TBA). After these seminal papers tremendous effort was devoted
to the calculation of correlation functions using various approaches
\cite{PhysRevD.23.3081,creamer-thacker-wilkinson-rossz,korepinBook,springerlink:10.1007/BF01029221,LL-master-eq-maillet-karol,Caux-Calabrese1,XXZ-to-LL}. One
of the most important recent results is the exact determination of the
long-distance behaviour of correlations \cite{karol-corr1,karol-corr2,adilet1,adilet-caux}.

Apart from purely academic interest, the study of the 1D
Bose gas was spurred by the recent success of experiments with cold
atoms in quasi one-dimensional traps
\cite{LL-exp1,LL-exp2,low-D-trapped,Druten-Yang-Yang,LL-exp3,g3-g4,LL-exp-g3}. A
remarkable result was presented in 
\cite{Druten-Yang-Yang}, where the authors managed to measure exact
predictions of the
TBA (for further developments and open questions see \cite{Druten-YangYang2}).
In experimental situations the local correlations are of special
interest, for example the three-body local correlation is related to
the rate of particle loss \cite{g3,LL-g3-exp-2,LL-exp-g3} and to the third moment
of the density fluctuations
\cite{LL-densityfluct-1,LL-densityfluct-2}. Moreover, even the
four-body 
correlations might be accessible to experiment, as it was recently demonstrated
in a 3D experiment \cite{g3-g4}. 

Concerning the general
$K$-body local
correlations (for the precise definition see the main text) there has been considerable
theoretical progress, too. The $K=1$ case is simply given by the
(linear) density of particles, whereas the $K=2$ case was related to the
thermodynamical quantities of the model in \cite{GaShly-gK-2}. Concerning the
higher-body cases small-coupling and
large-coupling expansions were performed in \cite{GaShly-gK-2,GaShly-gK-1}, 
whereas the exact  ground state value of the three-body correlation was
calculated in \cite{zvonarev-g3}. A new approach was initiated in
\cite{sinhG-LL1,sinhG-LL2}, where an infinite integral series (also
called the LeClair-Mussardo or LM series) was
derived using a special non-relativistic limit of the sinh-Gordon
model. The LM series applies for any $K$ and arbitrary temperature,
including the ground state, and it can be considered as an effective
large-coupling expansion of the quantity in question. The papers
\cite{nonrelFF,LM-sajat} considered the relation between the LM series
and previous form factor calculations with the Algebraic Bethe
Ansatz (ABA); in \cite{LM-sajat} it was shown that the LM series can be
understood and proven within the ABA. However, there was one crucial
problem: there were no explicit and general results
available for the form factors entering the LM series; the
numerical results in \cite{sinhG-LL1,sinhG-LL2} were obtained using a
truncation of the full series.

The important task of the exact summation of the LM series was
performed for the first time  in the recent article \cite{g3-exact},
where the authors evaluated the three-body correlation based on a
well-supported conjecture for the corresponding form factors. To our
best knowledge this is the first time that an exact, explicit and
compact result was given for a  non-trivial correlation
of the 1D Bose gas, valid at arbitrary couplings and temperatures.

In the present work we contribute to the calculation of the $K$-body
correlators using a different approach. Our strategy is the
following. First we consider a related physical quantity (the
so-called ``emptiness formation probability'') on a generic XXZ spin
chain and show that a special scaling limit of the spin chain
\cite{XXZ-to-LL,XXZ-LL-elso-emlites} yields the desired correlations
in the Bose gas (Section \ref{XXZ}). The matrix elements of the
operator on the spin chain are calculated in Section \ref{XXZ-FF}
borrowing results from the works
\cite{spin-spin-XXZ,maillet-inverse-scatt}. We then perform the
scaling limit towards the Bose gas in Section \ref{LL-FF}, this way we
obtain the form factors in a finite volume, with a finite number of
particles. Finally, the thermodynamic limit is performed in the Bose
gas (Section \ref{LL-FF-thermo}) leading to the multiple integral
\eqref{ordoK}, which is the 
main result of this work (see \eqref{gK} for the dimensionless form).

In principle the multiple integrals could be evaluated for any
$K$, but in practice this becomes more and more difficult with
increasing $K$, therefore it is desirable to derive more compact
results. 
In Section \ref{factorization} we show how the factorize the multiple
integral in the cases $K\le 4$. The results are the expressions \eqref{J2},
  \eqref{J3} and \eqref{J4}. In subsection \ref{factdimless} we also
  present examples of the numerical results.

Finally in Section \ref{LM} we determine all form factors entering
a modified form of the LM series, making it an explicit
integral series for the $K$-body local correlations.

\section{The Lieb Liniger model}

\label{LL}

The second quantized form of the Hamiltonian is
\begin{equation}
\label{H-LL}
H_{\text{LL}}=\int_{0}^{L}\,\mathrm d
x\left(\partial_x\Psi^\dagger\partial_x\Psi+ 
c\Psi^\dagger\Psi^\dagger\Psi\Psi\right).
\end{equation}
Here $L$ is the size of the system, periodic boundary conditions are
understood and
$\Psi(x,t)$ and $\Psi^\dagger(x,t)$ are  canonical
non-relativistic Bose
fields satisfying
\begin{equation}
  [\Psi(x,t),\Psi^\dagger(y,t)]=\delta(x-y).
\end{equation}
We used the conventions  $m=1/2$ and $\hbar=1$ and  $c>0$ is the
coupling constant. 

The eigenstates of the Hamiltonian \eqref{H-LL} can be constructed
using the Bethe Ansatz \cite{Lieb-Liniger,Lieb2,korepinBook}.
The $N$-particle coordinate space wave function is given by
\begin{equation}
\label{egyfajta-coo}
  \chi_N(p|x)=\frac{1}{\sqrt{N!}} \sum_\Pe 
\exp\left\{ i\sum_j x_j (\Pe p)_j \right\} \prod_{j>k}
\frac{(\Pe p)_j-(\Pe p)_k-ic\epsilon(x_j-x_k)}{(\Pe p)_j-(\Pe p)_k},
\end{equation}
where $\epsilon(x)$ is the sign function.

Periodic boundary conditions force the quasi-momenta to be solutions
of the Bethe Ansatz equations
\begin{equation}
  \label{LL-BA-e}
  e^{ip_jL}\prod_{k\ne j} \frac{p_j-p_k-ic}{p_j-p_k+ic}=1.
\end{equation}
The energy and momentum of the multi-particle state is given by
\begin{equation*}
  E_N=\sum_j p_j^2\qquad\qquad P_N=\sum_j p_j.
\end{equation*}
The norm of the wave function \eqref{egyfajta-coo} is \cite{gaudin-LL-norms,korepin-norms}
\begin{equation}
  \label{egyfajta-norm}
\mathcal{N}^{LL}= \int  \ |\chi_N|^2=\prod_{j<k} \frac{(p_j-p_k)^2+c^2}{(p_j-p_k)^2}\times
\det \mathcal{G}^{LL}
\end{equation}
with
\begin{equation}
  \mathcal{G}^{LL}_{jk}=\delta_{j,k}\left(
L+\sum_{l=1}^N \varphi(p_j-p_l)\right) -\varphi(p_j-p_l)
\end{equation}
and
\begin{equation*}
  \varphi(u)=\frac{2c}{u^2+c^2}.
\end{equation*}

We will be interested in the matrix elements of the operators
\begin{equation*}
  \ordo_K=\left(\Psi^\dagger(0)\right)^K \left(\Psi(0)\right)^K.
\end{equation*}
In coordinate space the matrix elements are given by the integrals
\begin{equation}
\label{eloiras}
\begin{split}
&  \bra{\phi_N}\ordo_K \ket{\chi_N}=\frac{N!}{K!(N-K)!}\times\\
&\hspace{1cm}\int_0^L dx_1 \dots dx_{N-K}\ \phi_N^*(0,\dots,0,x_1,\dots,x_{N-K})
\chi_N(0,\dots,0,x_1,\dots,x_{N-K}).
\end{split}
\end{equation}
The expectation value of $\ordo_K$ describes the probability to have
$K$ particles at the same point. It is useful to introduce the dimensionless 
quantities
\begin{equation*}
  g_K=\frac{\vev{\ordo_K}}{n^K},
\end{equation*}
where $n=N/L$ is the particle density. 
It can be shown by scaling arguments that in the thermodynamic limit
$g_K$ only depends on the dimensionless parameters
\begin{equation*}
  \gamma=\frac{c}{n}\qquad\qquad\tau=\frac{T}{n^2},
\end{equation*}
where $T$ is the temperature (we used the convention $k_B=1$
for the Boltzmann constant).

In principle the form factors \eqref{eloiras} could be obtained by
performing the integrals in coordinate space, but this becomes
increasingly complicated with growing $N$. Note also that the
Algebraic Bethe Ansatz for the Bose gas does not lead to simple
results either: the action of the field operators on Bethe states
 can be evaluated easily, but afterwards one would have to compute
scalar products of Bethe states with a reduced set of $N-K$ particles,
neither of which are on-shell, and there is no good formula for the
scalar products of such states. One way out of these problems is to consider
 a related quantity (the
``emptiness formation probability'') on 
the XXZ spin chain, where there are methods available to compute its
matrix elements. 

\section{The XXZ chain and its special scaling limit}

\label{XXZ}

The XXZ spin chain with $M$ sites and periodic boundary conditions  is
given by the following Hamiltonian: 
\begin{equation}
  \label{XXZ-H}
  H=J\sum_{j=1}^{M}
  (S^x_jS^x_{j+1}+S^y_jS^y_{j+1}+\Delta
(S^z_jS^z_{j+1}-1/4))+h\sum_{j=1}^M S_j^z.
\end{equation}
This model is also solvable by the Bethe Ansatz
\cite{XXX,XXZ1,XXZ2,XXZ3}. The $N$-particle eigenstates are given by 
\begin{equation}
\label{XXZ-eloallitas}
  \ket{\phi_N}=\frac{1}{\sqrt{N!}}
\sum_{y_1=1}^L \dots \sum_{y_1=N}^L
  \phi_N(\lambda|y_1,\dots,y_N)   \sigma^-_{y_1}\dots \sigma^-_{y_N} \ket{0}.
\end{equation}
Here $\ket{0}$ is the reference state with all spins up and $y_j$ are
the positions of the down spins.
The amplitudes are
\begin{equation}
\label{blue-vortex}
  \phi_N(\lambda|y)=\frac{1}{\sqrt{N!}}\sum_{\Pe \in\sigma_N} 
\prod_{1\le m < n \le N}
\frac{\sinh((\Pe\lambda)_{m}-(\Pe\lambda)_{n}+\epsilon(y_n-y_m)\eta)}
{\sinh((\Pe\lambda)_m-(\Pe\lambda)_n)}
\prod_{l=1}^N F((\Pe \lambda)_l,y_l),
\end{equation}
where
\begin{equation}
\label{inhom}
  F(\lambda,y)=\frac{1}{\sinh(\lambda-\xi_y)}\prod_{j=1}^{y-1} 
\frac{\sinh(\lambda-\xi_j+\eta)}{\sinh(\lambda-\xi_j)}.
\end{equation}
The parameter $\eta$ is related to the anisotropy:
\begin{equation*}
  \Delta=\cosh\eta.
\end{equation*}
In \eqref{inhom} we introduced inhomogeneities $\xi_j$ for the sites of
the spin chain; they will be used as a technical tool to obtain the
form factors in section \ref{XXZ-FF}. The physical limit consists of
setting all $\xi_j\to \eta/2$. The expression \eqref{blue-vortex} is a
seemingly over-complicated way to write down the wave function,
because it is valid at arbitrary values of the variables $y_j$ and not
only in the region $y_1<\dots<y_M$. We used this form to have an exact
agreement with the conventions used in \eqref{egyfajta-coo}.

The Bethe equations follow from the periodicity of the wave function
and they read
\begin{equation}
  \label{BAe}
d(\lambda_j)
\prod_{k\ne j}
\frac{\sinh(\lambda_j-\lambda_k+\eta)}{\sinh(\lambda_j-\lambda_k-\eta)}=1,
\end{equation}
where
\begin{equation}
\label{illuminus}
  d(\lambda)=\prod_{k=1}^M
  \frac{\sinh(\lambda-\xi_k)}{\sinh(\lambda-\xi_k+\eta)}.
\end{equation}
In the normalization \eqref{XXZ-eloallitas}-\eqref{blue-vortex}  the
norm of the wave function is given by 
\begin{equation}
\label{XXZ-norm}
\begin{split}
&\mathcal{N}^{XXZ}=\sum_{y_1}\dots \sum_{y_N}\ |\phi(y_1,\dots,y_N)|^2=\\
&\hspace{4cm}(-\sinh\eta)^{-N} \prod_{j<k} f(\lambda_j,\lambda_k) f(\lambda_k,\lambda_j)
\times\det \mathcal{G}^{XXZ}
\end{split}
\end{equation}
with
\begin{equation}
  \mathcal{G}^{XXZ}_{jk}=\delta_{j,k}\left(
\frac{d'(\lambda_j)}{d(\lambda_j)}+\sum_{l\ne j}\varphi^{XXZ}(\lambda_j-\lambda_l)
\right)
-\varphi^{XXZ}(\lambda_j-\lambda_k).
\end{equation}
The kernel $\varphi^{XXZ}$ is given by
\begin{equation}
  \varphi^{XXZ}(u)=\frac{-\sinh\eta}{\sinh(u+\eta/2)\sinh(u-\eta/2)}.
\end{equation}
One-particle momenta and energies are given by the formulas
\begin{equation*}
  e^{ip(\lambda)}=\frac{\sinh(\lambda+\eta/2)}{\sinh(\lambda-\eta/2)}\qquad
\qquad
e(\lambda)=J\frac{\sinh^2\eta}{\cos(2\lambda)-\cosh\eta}-h.
\end{equation*}

\subsection{Towards the Lieb-Liniger model}

There is a special scaling limit of the XXZ chain which yields the
physical quantities of the Lieb-Liniger model
\cite{Perk-XXZ-to-LL,XXZ-LL-elso-emlites,XXZ-to-LL}. In order to obtain the Bose gas
in a finite volume $L$ one has to set
\begin{equation*}
  \eta=i\pi-i\eps\qquad M=\frac{c}{\eps^2}L
\end{equation*}
and let $\eps\to 0$ (here $c$ is the
coupling constant of the Bose gas). The number $N$ of the magnons has
to be kept fixed and the rapidities of the particles have to be scaled as
\begin{equation*}
\lambda_j=p_j \frac{\eps}{c}.
\end{equation*}
After the limiting procedure the magnons can be identified as the particles of the
Bose gas with rapidity $p_j$.
It can be shown that under an appropriate scaling of the parameters $J$
and $h$  
\begin{equation*}
  e(\lambda)=J\frac{\sinh^2\eta}{\cos(2\lambda)-\cosh\eta}-h
\qquad\to\qquad {p^2}-\mu,
\end{equation*}
where $\mu$ is the chemical potential in the Bose gas. However, this
will be not needed in the following; we will consider
 the Bethe wave functions
and the form factors of local operators. In the following we 
assume that the homogeneous limit $\xi_j\to\eta/2$
is performed first  on the spin chain, and the limit towards the Bose gas is taken
afterwards. 

Taking the scaling limit of the Bethe equations \eqref{BAe} results in
\begin{equation}
  \label{BAelim}
(-1)^M   e^{i\nu_j l} \prod_{k\ne j} \frac{\nu_j-\nu_k-ic}{\nu_j-\nu_k+ic}=1.
\end{equation}
For the sake of simplicity we only consider even
chains so that no twist appears in the Bethe equations.

The limiting form of the Bethe wave function can be taken by setting 
$x_j=\frac{\eps^2}{c} y_j$ and keeping $x_j$ finite, which will correspond
to the position of the particles of the Bose gas. The wave function then reads
\begin{equation}
\label{psylocibe}
  \Psi_N(x|\nu)=\frac{1}{\sqrt{N!}}\sum_{\Pe\in\sigma_N} 
\prod_{ m > n } -\frac{(\Pe\nu)_m-(\Pe\nu)_n+i\epsilon(x_m-x_n)c}{(\Pe\nu)_m-(\Pe\nu)_n}
\prod_{l=1}^N F((P\nu)_l,x_l),
\end{equation}
where
\begin{equation}
  F(\nu,x)=
e^{-i\nu x} (-1)^{y}.
\end{equation}
Apart from factors of $(-1)$ the above expression is  equal to the
complex conjugate of the Bethe wave function \eqref{egyfajta-coo}. It
can be argued that the
factors of $(-1)^{y_j}$ 
don't affect the calculation of form factors of local operators. Indeed, for any
coordinate space calculation one has to take the product of two wave
functions with the down spins placed at prescribed
positions. Depending on the operator in question an overall factor of
$(-1)$ may remain, but the position dependent factors of $(-1)^{y_j}$
always cancel. For the operators considered in this paper every such
factor cancels, therefore they will be neglected in the following.

Due to the relation between $y$ and $x$ it is expected that the norm
of the wave function behaves as
\begin{equation}
   \mathcal{N}^{XXZ}\quad\to\quad 
\left(\frac{c}{\eps^2}\right)^N
\mathcal{N}^{LL}.
\end{equation}
Comparing the formulas \eqref{egyfajta-norm} and \eqref{XXZ-norm} we
obtain the same scaling: the Gaudin determinants behave as
\begin{equation*}
  \det \mathcal{G}^{XXZ} \quad\to\quad 
\left(\frac{ic}{\eps}\right)^N  \det \mathcal{G}^{LL} 
\end{equation*}
and the prefactors contribute an extra $(i\eps)^{-N}$.

\subsection{The emptiness formation probability}

We are interested in the local operators $E_j^{\alpha\beta}$
acting on site $j$ with matrix elements 
\begin{equation*}
\Big(  E^{\alpha\beta}\Big)_{kl}=\delta_{k,\alpha}\delta_{l,\beta}.
\end{equation*}
In particular we consider the composite operator
\begin{equation}
  s_K=E_1^{--}E_2^{--}\dots E_K^{--}.
\end{equation}
When sandwiched between two states, this operator forces $K$
particles to occupy the first $K$ sites. The expectation value of
$s_K$ (or sometimes its spin reverse) is called the ``emptiness
formation probability''. 

We will show
 that the operator $s_K$ scales to
$\ordo_K$ in the limiting procedure. 
In the coordinate Bethe Ansatz its $N$-particle form factors  are given by
\begin{equation}
  \label{u-recken}
\begin{split}
&\bra{\{\lambda\}}s_K\ket{\{\mu\}}=\frac{N!}{K!(N-K)!}\times \\
&\sum_{y_1,\dots,y_{N-K}=K+1}^M\
\phi^*_N(\lambda|1,2,\dots,K,y_1,\dots,y_{N-K})\phi_N(\mu|1,2,\dots,K,y_1,\dots,y_{N-K}).
\end{split}
\end{equation}
This formula has to be compared to \eqref{eloiras}. It is easy to see
that the scaling limit of the Bethe wave functions works even if
a fixed number of particles are placed on the first few sites:
\begin{equation*}
  \phi_N(\mu|1,2,\dots,K,y_1,\dots,y_{N-K}) \quad\to\quad
\chi_N(p|0,0,\dots,0,x_1,\dots,x_{N-K})^*.
\end{equation*}
Therefore the un-normalized form factor will behave as
\begin{equation*}
\left(\frac{\eps^2}{c}\right)^{M-K} \bra{\{\lambda\}}s_K\ket{\{\mu\}}
\quad\to\quad
\bra{\{p\}}\ordo_K\ket{\{k\}}^*,
\end{equation*}
where it is understood that
\begin{equation*}
\frac{c}{\eps}\lambda_j\to p_j \quad\qquad
\frac{c}{\eps}\mu_j\to k_j.
\end{equation*}
For the normalized form factors we get
\begin{equation}
\label{normalis}
\left(\frac{\eps^2}{c}\right)^{-K} 
\frac{\bra{\{\lambda\}}s_K\ket{\{\mu\}}}{\sqrt{\skalarszorzat{\lambda}{\lambda}\skalarszorzat{\mu}{\mu}}}
\quad\to\quad
\frac{\bra{\{p\}}\ordo_K\ket{\{k\}}^*}{\sqrt{\skalarszorzat{p}{p}\skalarszorzat{k}{k}}}.
\end{equation}
Our strategy is to obtain explicit determinant formulas for the matrix
elements \eqref{u-recken} and to take the scaling limit according to \eqref{normalis}.

\section{Form factors in the XXZ chain}

\label{XXZ-FF}
 
In this section we compute explicit determinant formulas for the
matrix elements \eqref{u-recken} in the framework of Algebraic Bethe
Ansatz (ABA). Mostly we will use the results of
the papers 
\cite{spin-spin-XXZ,maillet-inverse-scatt};
the only difference between the present approach and the traditional
methods is that here the 
homogeneous limit $\xi_j\to\eta/2$ is taken explicitly before performing the thermodynamic
limit or the limit towards the Bose gas. 

The central object in ABA is the monodromy matrix, a $2\times 2$
matrix in the so-called auxiliary space with
operator valued entries which act on the Hilbert space of the spin
chain:
\begin{equation*}
  T(u)=
  \begin{pmatrix}
    A(u) & B(u) \\ C(u) & D(u)
  \end{pmatrix}.
\end{equation*}
It is built from the so-called local L-matrices:
\begin{equation*}
  T(u)=L_M(u)\dots L_1(u),
\end{equation*}
where
\begin{equation*}
  L_j(u)=R_{0j}(u-\xi_j).
\end{equation*}
Here  $j$ refers to the quantum space of the spin at site $j$
and 0 refers to the auxiliary space and the parameters $\xi_j$ are
identical to the inhomogeneities already introduced in \eqref{inhom}.
The operator $R(u)$ is the
R-matrix of the XXZ type:
\begin{equation}
  R(u)=\frac{1}{\sinh(u+\eta)}
  \begin{pmatrix}
    \sinh(u+\eta) & & &\\
& \sinh(u)  & \sinh(\eta) & \\
& \sinh(\eta) & \sinh(u) & \\
& & & \sinh(u+\eta)
  \end{pmatrix}.
\label{R}
\end{equation}
The trace of the monodromy matrix is called the transfer matrix:
\begin{equation*}
  \tau(u)=A(u)+D(u).
\end{equation*}
In the homogeneous limit $\xi_j=\eta/2$ it is related to the
Hamiltonian  \eqref{XXZ-H} at $h=0$ as
\begin{equation*}
  H\quad \sim \quad \frac{d}{du} \tau(u)\Big|_{u=\eta/2}+\text{const.}
\end{equation*}
The normalization \eqref{R} of the $R$-matrix results in the following vacuum
eigenvalues:
\begin{equation*}
  A(u)\ket{0}=\ket{0}\qquad D(u)\ket{0}=d(u)\ket{0},
\end{equation*}
with $d(\lambda)$ given by \eqref{illuminus}.

In the framework of ABA the Bethe states are
\begin{equation*}
\bra{0}\prod_{j=1}^N C(\lambda_j)
\quad\text{and}\quad  \prod_{j=1}^N B(\mu_j) \ket{0}.
\end{equation*}
They are eigenstates of the transfer matrix if the rapidities satisfy
the Bethe equations \eqref{BAe}. Apart from an overall normalization
factor  they are identical to the states given by the
coordinate wave functions \eqref{XXZ-eloallitas}.

In order to compute the matrix elements of $s_K$ in the framework of
ABA the local operators have to be expressed in terms of the entries
of the transfer matrix. This problem was solved in
 \cite{maillet-inverse-scatt,goehmann-korepin-inverse} leading to the
 following theorem:
\begin{equation}
  \label{inverse}
E_j^{\alpha\beta}=\prod_{k=1}^{j-1}(A+D)(\xi_k)\times T^{\alpha\beta}(\xi_j)
\times \prod_{k=j+1}^M (A+D)(\xi_k).
\end{equation}
Applying this formula to operators $E^{--}_j$ on neighbouring sites one gets
\begin{equation}
  \label{carbon-based-lifeforms}
  s_k=D(\xi_1)D(\xi_2)\dots D(\xi_k) \prod_{l=k+1}^L  (A+D)(\xi_l).
\end{equation}
Evaluated on Bethe states
equation \eqref{carbon-based-lifeforms} yields
\begin{equation}
  \label{carbon-based-lifeforms2}
 \bra{0}\prod_{j=1}^N C(\lambda_j)\ s_k\
\prod_{j=1}^N B(\mu_j) \ket{0}
=\bra{0}\prod_{j=1}^N C(\lambda_j) \ D(\xi_1)D(\xi_2)\dots D(\xi_k) \
\prod_{j=1}^N B(\mu_j) \ket{0}\times
\prod_{j=1}^k \frac{1}{t(\xi_j,\{\mu\})}
\end{equation}
with $t(u,\{\mu\})$ being the corresponding eigenvalue of the transfer matrix. Evaluated
at the inhomogeneities it reads
\begin{equation}
  t(\xi_j,\{\mu\})=\prod_{k=1}^N \frac{\sinh(\mu_k-\xi_j+\eta)}{\sinh(\mu_k-\xi_j)}.
\end{equation}
In \eqref{carbon-based-lifeforms2} we also used the fact that
\begin{equation*}
   \prod_{l=1}^L  (A+D)(\xi_l)=1.
\end{equation*}
The action of multiple $D$ operators on the dual state results in \cite{spin-spin-XXZ} 
\begin{equation*}
  \label{fasza}
\begin{split}
&\bra{0}\prod_{j=1}^N C(\lambda_j) D(\xi_1)D(\xi_2)\dots D(\xi_k) 
 =\\
&\mathop{\sum_{\{\lambda^+\}\cup \{\lambda^-\}}}_{|\{\lambda^+\}|=k} 
\frac{\prod_{o,p} \sinh(\lambda_p^+-\xi_o+\eta)}
{\prod_{j<l}\sinh(\xi_l-\xi_j)\sinh(\lambda^+_j-\lambda^+_l)}\times
\det t(\lambda^+_l,\xi_j) \times
\prod_{o,p} f(\lambda^+_p,\lambda^-_o)\times\\
&\hspace{4cm}
\prod_l d(\lambda_l^+)\bra{0}\prod_o C(\lambda^-_o)C(\xi_1)\dots C(\xi_k)
\end{split}
\end{equation*}
with
\begin{equation*}
  t(\lambda,\xi)=\frac{\sinh\eta}{\sinh(\lambda-\xi)\sinh(\lambda-\xi+\eta)}.
\end{equation*}
The scalar product of an arbitrary state and a Bethe state is \cite{slavnov-overlaps}
\begin{equation}
  \label{scalar}
\bra{0}\prod_j C(\lambda_j)\prod_j B(\mu_j)  \ket{0}=
\frac{\prod_{j,k}\sinh(\mu_j-\lambda_k+\eta)}{\prod_{j<k}
  \sinh(\lambda_j-\lambda_k)\sinh(\mu_k-\mu_j)} \times \det S,
\end{equation}
where
\begin{equation*}
  S_{jk}=t(\mu_j,\lambda_k)-d(\lambda_k)t(\lambda_k,\mu_j)
\prod_{l=1}^N \frac{\sinh(\lambda_k-\mu_l+\eta)}{\sinh(\lambda_k-\mu_l-\eta)}.
\end{equation*}
Specializing this to the present case
\begin{equation*}
\begin{split}
&\bra{0}\prod_o C(\lambda_o^-)  C(\xi_1)\dots C(\xi_k) \prod_j
B(\mu_j)\ket{0}=\\
&\frac{\prod_{j,k}\sinh(\mu_j-\lambda_k^-+\eta)\prod_{j,o}\sinh(\mu_j-\xi_o+\eta)}
{\prod_{j<k}\sinh(\mu_k-\mu_j)\prod_{j<k}\sinh(\lambda_j^--\lambda_k^-)
\prod_{j<k}\sinh(\xi_j-\xi_k)\prod_{j,k}\sinh(\xi_j-\lambda_k)
}\times\\
&\times \det U 
.
\end{split}
\end{equation*}
Here
\begin{equation}
\begin{split}
\label{Uuu}
  U_{jl}&=t(\mu_j,\xi_l) \qquad\text{for}\quad l=1\dots K\\
U_{j,k+l}&=t(\mu_j,\lambda_l^-)-d(\lambda_l^-)t(\lambda_l^-,\mu_j)
\prod_{o=1}^N
\frac{\sinh(\lambda_l^--\mu_o+\eta)}{\sinh(\lambda_l^--\mu_o-\eta)}
\qquad\text{otherwise.}
\end{split}
\end{equation}
Therefore, the form factor of the inhomogeneous chain is given by
\begin{equation}
  \label{larger-than-life}
\begin{split}
&\bra{\{\lambda\}}s_k\ket{\{\mu\}}=\\
&\prod_{j=1}^k \frac{1}{t(\xi_j)}\times
\mathop{\sum_{\{\lambda^+\}\cup \{\lambda^-\}}}_{|\{\lambda^+\}|=k} 
\frac{\prod_{o,p} \sinh(\lambda_p^+-\xi_o+\eta)}
{\prod_{j<l}\sinh(\xi_l-\xi_j)\sinh(\lambda^+_j-\lambda^+_l)}
\prod_{o,p} f(\lambda^+_p,\lambda^-_o)\times \prod_l d(\lambda_l^+)\\
&\frac{\prod_{j,k}\sinh(\mu_j-\lambda_k^-+\eta)\prod_{j,o}\sinh(\mu_j-\xi_o+\eta)}
{\prod_{j<k}\sinh(\mu_k-\mu_j)\prod_{j<k}\sinh(\lambda_j^--\lambda_k^-)
\prod_{j<k}\sinh(\xi_j-\xi_k)\prod_{j,k}\sinh(\xi_j-\lambda_k)
}\times\\
&\times \det t(\lambda^+_l,\xi_j) \times \det U.
\end{split}
\end{equation}

We now perform the homogeneous limit $\xi_j\to\eta/2$ following the method of
\cite{six-vertex-partition-function}. For the matrix $M$ we get
\begin{equation*}
  \lim \frac{\det
    M}{\prod_{j>l}\sinh(\xi_j-\xi_k)}=\frac{1}{\prod_{\alpha=1}^{m-1} \alpha!}
\det H
\end{equation*}
with
\begin{equation*}
  H_{jl}=\left[\left(\frac{\partial}{\partial \xi}\right)^{l-1} t(\lambda_k,\xi)\right]|_{\xi=\eta/2}.
\end{equation*}
It is advantageous to use the form
\begin{equation*}
  t(\lambda,\xi)=\frac{\cosh(\lambda-\xi)}{\sinh(\lambda-\xi)}-
\frac{\cosh(\lambda-\xi+\eta)}{\sinh(\lambda-\xi+\eta)}.
\end{equation*}
Taking the derivatives one is free to replace \cite{2010NJPh...12e5028M}
\begin{equation*}
  \left(\frac{\partial}{\partial \xi}\right)^{l-1} t(\lambda,\xi)\quad
\to\quad
(-1)^{l-1}(l-1)!\left[\left( \frac{\cosh(\lambda-\xi)}{\sinh(\lambda-\xi)}\right)^{l}-
\left(\frac{\cosh(\lambda-\xi+\eta)}{\sinh(\lambda-\xi+\eta)}\right)^{l}\right].
\end{equation*}
The same steps can be performed for the corresponding elements of 
the matrix $U$. Finally the homogeneous limit reads
\begin{equation}
  \label{you-didnt-believe-it-was-going-to-be-that-easy-did-you}
\begin{split}
&\bra{\{\lambda\}}s_K\ket{\{\mu\}}=\mathop{\sum_{\{\lambda^+\}\cup \{\lambda^-\}}}_{|\{\lambda^+\}|=K} 
\frac{\prod_{o,p} \sinh(\lambda_p^++\eta/2)}
{\prod_{j<l}\sinh(\lambda^+_j-\lambda^+_l)}
\prod_{o,p} f(\lambda^+_p,\lambda^-_o)\times \prod_l d(\lambda_l^+)\times\\
&\frac{\prod_{j,k}\sinh(\mu_j-\lambda_k^-+\eta)\prod_{j}\sinh^K(\mu_j-\eta/2)}
{\prod_{j<k}\sinh(\mu_j-\mu_k)\prod_{j<k}\sinh(\lambda_j^--\lambda_k^-)
\prod_l\sinh^K(\eta/2-\lambda_l)}\times \det O \det V.
\end{split}  
\end{equation}
with
\begin{equation}
  \label{OO}
O_{jl}=
\left[\left( \frac{\cosh(\lambda_j^+-\eta/2)}{\sinh(\lambda_j^+-\eta/2)}\right)^{l}-
\left(\frac{\cosh(\lambda_j^++\eta/2)}{\sinh(\lambda_j^++\eta/2)}\right)^{l}\right]
\end{equation}
and
\begin{equation}
\begin{split}
\label{UuuV}
  V_{jl}&=
\left[\left( \frac{\cosh(\mu_j-\eta/2)}{\sinh(\mu_j-\eta/2)}\right)^{l}-
\left(\frac{\cosh(\mu_j+\eta/2)}{\sinh(\mu_j+\eta/2)}\right)^{l}\right]
 \qquad\text{for}\quad l=1\dots K\\
V_{j,K+l}&=t(\mu_j,\lambda_l^-)-d(\lambda_l^-)t(\lambda_l^-,\mu_j)
\prod_{o=1}^N
\frac{\sinh(\lambda_l^--\mu_o+\eta)}{\sinh(\lambda_l^--\mu_o-\eta)}
\qquad\text{otherwise.}
\end{split}
\end{equation}

\section{The scaling limit of the form factors}

\label{LL-FF}

Here we take the scaling limit of the formula
\eqref{you-didnt-believe-it-was-going-to-be-that-easy-did-you} to
obtain the matrix elements of $\ordo_K$ in the Bose gas. We substitute
\begin{equation*}
  \eta=i\pi-i\eps\quad \quad\lambda=\frac{\eps}{c}p\quad\quad \mu=\frac{\eps}{c}k.
\end{equation*}
It is straightforward to calculate the limiting values of the prefactors, but
the determinants need special care. The elements of $O$ read
\begin{equation}
  \label{O2}
O_{jl}=
\left[\left( \frac{\sinh(\lambda_j^++i\eps/2)}{\cosh(\lambda_j^++i\eps/2)}\right)^{l}-
\left(\frac{\sinh(\lambda_j^+-i\eps/2)}{\cosh(\lambda_j^+-i\eps/2)}\right)^{l}\right].
\end{equation}
The leading terms will be
\begin{equation*}
  O_{jl}
\quad\to\quad l   \left(\frac{\eps}{c}\right)^{l-1} \left(p_j^+\right)^{l-1} i\eps,
\end{equation*}
which yields
\begin{equation*}
  \det O\quad\to\quad K! (i\eps)^K \left(\frac{\eps}{c}\right)^{(K-1)K/2} \det
 \left[ (p_j^+)^{l-1}\right]=
K! (i\eps)^K  \left(\frac{\eps}{c}\right)^{(k-1)k/2} \prod_{j>l} (p_j^+-p_l^+).
\end{equation*}
One can use the same expansion for the first $K$ columns of the matrix $V$.

To obtain the proper normalization note that in ABA the norm of the Bethe state scales as
\begin{equation*}
  \bra{0}\prod_j C(\lambda_j)\prod_j B(\lambda_j)  \ket{0}\quad
\to\quad
 c^N \frac{\prod_{j< l}
 ( (p_j-p_l)^2+c^2)}{\prod_{j<l}(p_j-p_l)^2 }\times \det \mathcal{G}_{LL}.
\end{equation*}
This differs from \eqref{egyfajta-norm} by the overall factor of
$c^N$. 

Collecting all the factors and performing a complex conjugation we find
\begin{equation}
  \label{haaatC}
  \begin{split}
&\bra{\{p\}}\ordo_K\ket{\{k\}}=
c^{K-N} 
(K!)^2
\mathop{\sum_{\{p^+\}\cup \{p^-\}}}_{|\{p^+\}|=K} 
\prod_{o,l}  f(p^+_l,p^-_o) \prod_l e^{-iLp_l^+}\times  \\
&\hspace{6cm}\times\frac{\prod_{j,l} (k_j-p_l^-+ic)}
{\prod_{j<l}(k_j-k_l)\prod_{j<k}(p_j^--p_k^-)}
\times \det Z,
\end{split}  
\end{equation}
with
\begin{equation}
  \begin{split}
    Z_{jl}&=(k_j)^{l-1}
 \qquad\text{for}\quad l=1\dots K\\
 Z_{j,K+l}&= t(k_j,p_l^-)-e^{-iLp_l^-} t(p_l^-,k_j)
\prod_{o=1}^N
\frac{(p_l^--k_o+ic)}{(p_l^--k_o-ic)}
\qquad\text{otherwise.}
  \end{split}
\end{equation}
Here we used
\begin{equation*}
 t(u)=\frac{ic}{u(u+ic)}.
\end{equation*}
 Equation \eqref{haaatC} refers to the normalization where the norms
of the states $\ket{\{p\}}$ and $\ket{\{k\}}$ are given by \eqref{egyfajta-norm}.

The result \eqref{haaatC} is valid whenever the set $\{k\}$ satisfies
the Bethe equations. 
In the case when $\{p\}$ is also a solution but different from $\{k\}$ 
we obtain the form factors
\begin{equation}
  \label{FFK}
  \begin{split}
&    F^K_N(\{p\},\{k\})=c^{K-N} 
(K!)^2 
\mathop{\sum_{\{p^+\}\cup \{p^-\}}}_{|\{p^+\}|=K} 
\prod_{o,l}  f(p^-_o,p^+_l) \prod_l \times  \\
&\hspace{6cm}\times\frac{\prod_{j,l} (k_j-p_l^-+ic)}
{\prod_{j<l}(k_j-k_l)\prod_{j<k}(p_j^--p_k^-)}
\times \det V,
  \end{split}
\end{equation}
with
\begin{equation}
  \begin{split}
    V_{jl}&=(k_j)^{l-1}
 \qquad\text{for}\quad l=1\dots K\\
 V_{j,K+l}&= t(k_j,p_l^-)+ t(p_l^-,k_j)
\prod_{o=1}^N
\frac{(p_l^--k_o+ic)(p_l^--p_o-ic)}{(p_l^--k_o-ic)(p_l^--p_o+ic)}
\qquad\text{otherwise.}
  \end{split}
\end{equation}
This is a new result of the present work. 
In the case of $K=1$ eq. \eqref{FFK} yields an alternative representation
for the form factors of the density operator, which were previously
determined in \cite{slavnov-overlaps,springerlink:10.1007/BF01029221,korepin-slavnov}.

To obtain the mean value of $\ordo_K$ we take the limit
$\{p\}\to\{k\}$ in \eqref{haaatC} and divide by the norm
\eqref{egyfajta-norm} resulting in
\begin{equation}
\label{hat-ez-mar-a-sokadik-tipp}
\vev{\ordo_K}_N=(K!)^2
\mathop{\sum_{\{p^+\}\cup \{p^-\}}}_{|\{p^+\}|=K} 
\left[\prod_{j>l}\frac{p_j^+-p_l^+}{(p_j^+-p_l^+)^2+c^2)}\right]
\times \frac{\det \mathcal{H} }{\det \mathcal{G}^{LL}}.
\end{equation}
The elements of $\mathcal{H}$ are given by 
\begin{equation}
\label{maia}
 \mathcal{H}_{j,l}=
 \begin{cases}
(p_j)^{l-1}
&\text{for}\quad l=1\dots K\\
\mathcal{G}^{LL}_{j,l}&\text{for} \quad l=K+1\dots N.
 \end{cases}
\end{equation}
Here it is understood that in both $\mathcal{G}^{LL}$ and $\mathcal{H}$ the ordering of
the rapidities is given by $\{p\}=\{\{p^+\},\{p^-\}\}$. The
matrix $\mathcal{H}$ differs from $\mathcal{G}^{LL}$ only in those
columns which belong to the subset $\{p^+\}$.

In the case of $K=1$ the above formula results in
\begin{equation*}
  \vev{\ordo_1}=\frac{N}{L}
\end{equation*}
as it should. To prove this note that the sums of the columns of
$\mathcal{G}^{LL}$ are equal to $L$ in every row, therefore every
$\mathcal{H}$ gives
\begin{equation*}
  \mathcal{H}=\frac{1}{L} \mathcal{G}^{LL}.
\end{equation*}

\section{Expectation values in the thermodynamic limit}

\label{LL-FF-thermo}

In this section we evaluate the thermodynamic limit of
\eqref{hat-ez-mar-a-sokadik-tipp}.
We consider a Bethe state $\ket{\Omega}$ in a large volume $L$ with a
large number of particles such that the particle density $n=N/L$ is
fixed. In the thermodynamic limit one defines the density of roots
$\rho^{(r)}(p)$ and holes $\rho^{(h)}(p)$ and the total density
$\rho(p)=\rho^{(r)}(p)+\rho^{(h)}(p)$. This latter function satisfies
the Lieb-equation
\begin{equation}
  \rho(p)=\frac{1}{2\pi} +\int_{-\infty}^\infty dq\ \varphi(p-q)f(q)\rho(q).
\end{equation}
Here
\begin{equation}
\label{fp}
  f(p)=\frac{\rho^{(r)}(p)}{\rho(p)}
\end{equation}
is a distribution function characterizing the state in question.
In thermal equilibrium $f(p)=(1+e^{\eps(p)})^{-1}$ where $\eps(p)$ is
the so-called pseudo-energy, which is a solution of the TBA equation
\begin{equation}
\label{TBA}
   \eps(p)=\frac{p^2-\mu}{T}-
  \int_{-\infty}^\infty\frac{dp'}{2\pi} 
\varphi(p-p') \log(1+e^{-\eps(p')}).
\end{equation}  
At zero temperature we recover the ground-state distribution
\begin{equation}
\label{fpp}
  f(p)=
  \begin{cases}
    1 & |p|\le  \Lambda\\
0 & |p|>\Lambda
  \end{cases}
\end{equation}
with $\Lambda$ being the Fermi-rapidity. The particle number is always
given by the formula
\begin{equation*}
  n=\frac{N}{L}=\int f(p) \rho(p).
\end{equation*}

We proceed to calculate the thermodynamic limit of
\eqref{hat-ez-mar-a-sokadik-tipp} using the techniques of
\cite{KitanineMailletTerras-XXZ-corr1}.  
The ratio of determinants is
calculated as
\begin{equation*}
   \frac{\det \mathcal{H}}{\det \mathcal{G}^{LL}}=\det
\Big(   (\mathcal{G}^{LL})^{-1} \mathcal{H}\Big).
\end{equation*}
The resulting matrix on the r.h.s. will be equal to the identity
matrix except for those columns belonging to the set $\{p^+\}$. These
elements can be evaluated by transforming the action of
$\mathcal{G}^{LL}$ into an integral equation. 
This results in
\begin{equation*}
  \frac{\det \mathcal{H}}{\det \mathcal{G}^{LL}}=
\prod_o \frac{1}{2\pi L \rho(p_o^+)}\times \det I,
\end{equation*}
where
\begin{equation*}
  I_{jl}=h^{(l-1)}(p^+_j).
\end{equation*}
Here $h^{(l)}(u)$ is the solution of the linear integral equation
\begin{equation}
\label{hdef}
  h^{(l)}(p)=p^l +\int_{-\infty}^\infty \frac{dq}{2\pi}\ \varphi(p-q)f(q)h^{(l)}(q).
\end{equation}
Note that in this normalization $h^{(0)}(p)=2\pi \rho(p)$ and
$f(p)h^{(0)}(p)=2\pi \rho^{(r)}(p)$.

As a final step one integrates over the rapidities
$p^+_j$ and in 
 the thermodynamic limit one gets
\begin{equation}
\begin{split}
\label{this-is-the-end-beautiful-friend}
\bra{\Omega}\ordo_K\ket{\Omega}=
K!
 \int \frac{dp_1}{2\pi}\dots \frac{d p_K}{2\pi}
\prod_{o} f(p_o)   \prod_{j>l}\frac{p_j-p_l}{(p_j-p_l)^2+c^2}\times
 \det I.
\end{split}
\end{equation}
The prefactors are completely anti-symmetric therefore one can expand the
determinant and write
\begin{equation}
\label{ordoK}
\bra{\Omega}\ordo_K\ket{\Omega}=
(K!)^2 \int \frac{dp_1}{2\pi}\dots \frac{d p_K}{2\pi}
  \prod_{j>l}\frac{p_j-p_l}{(p_j-p_l)^2+c^2}
\prod_{o=1}^K f(p_o) h^{(o-1)}(p_o).
\end{equation}
This formula is the main result of our paper. Its explicit
factorization is performed in the cases $K=2,3,4$ in section \ref{factorization}.

For practical purposes it is useful to derive the dimensionless
multiple integral for the quantity 
\begin{equation*}
  g_K=\frac{\vev{\ordo_K}}{n^K},
\end{equation*}
which only depends on the dimensionless coupling constant $\gamma=c/n$ and
the dimensionless version of the distribution functions $f(p)$. We
define
\begin{equation*}
  q=\frac{p}{c}\qquad f(q)=f(p=qc).
\end{equation*}
In the finite temperature case $f(q)=(1+e^{\tilde\eps(q)})^{-1}$ where
$\tilde\eps(q)$ is the solution of the dimensionless equation
\begin{equation}
  \label{dimlessTBA}
   \tilde  \eps(q)=-\alpha +\frac{q^2\gamma^2}{\tau}-
  \int_{-\infty}^\infty\frac{dq'}{2\pi} 
\frac{2}{(q-q')^2+1)} \log(1+e^{-\eps(p')}),
\end{equation}
with
\begin{equation*}
  \alpha=\frac{\mu}{T}\qquad\qquad \tau=\frac{T}{n^2}.
\end{equation*}
Defining the dimensionless functions $\tilde h^{(l)}(q)$ as
\begin{equation}
\label{qdef}
 \tilde  h^{(l)}(q)=q^l +\int_{-\infty}^\infty \frac{dq'}{2\pi}\
 \frac{2}{(q'-q)^2+1}f(q')\tilde h^{(l)}(q')
\end{equation}
we find $\tilde h^{(l)}(q)=c^l h^{(l)}(p=qc)$. Thus the dimensionless
multiple integral formula is expressed as
\begin{equation}
\label{gK}
g_K={(K!)^2}\gamma^K \int \frac{dq_1}{2\pi}\dots \frac{d q_K}{2\pi}
  \prod_{j>l}\frac{q_j-q_l}{(q_j-q_l)^2+1}
\prod_{j=1}^K f(q_j) \tilde h^{(j-1)}(q_j).
\end{equation}

\subsection{The $c\to 0^+$ limit}

We take the small-coupling limit of the dimensionful formula
\eqref{this-is-the-end-beautiful-friend} by sending $c\to 0$ and
keeping $n$ fixed. The limiting form of the kernel $\varphi(u)$ is given by
\begin{equation}
\label{hukk}
  \varphi(u)\quad\to\quad 2\pi \delta(u).
\end{equation}
Therefore the solution of the integral equations \eqref{hdef} is
\begin{equation*}
  h^{(l)}(p)=\frac{p^l}{1-f(p)}.
\end{equation*}
The determinant in \eqref{this-is-the-end-beautiful-friend} has
the limiting value
\begin{equation*}
  \det I\quad\to\quad \prod_j \frac{1}{1-f(p)}\prod_{j>k} (p_j-p_k)
\end{equation*}
resulting in
\begin{equation*}
  \vev{\ordo_K}\quad\to\quad K!\prod_{j=1}^K
  \left[\int\frac{dp_j}{2\pi} \frac{f(p_j)}{1-f(p_j)} \right] .
\end{equation*}
Note that 
\begin{equation*}
\vev{\ordo_1}=n=  \int\frac{dp}{2\pi} \frac{f(p)}{1-f(p)},
\end{equation*}
therefore 
\begin{equation*}
g_K=\frac{ \vev{\ordo_K}}{n^K}\quad\to\quad K!
\end{equation*}
as it should be for free bosons.

The above calculation only applies if $f(p)<1$, therefore
the result is not valid for the ground state. In fact
\begin{equation*}
  \lim_{c\to 0}\lim_{T\to 0} g_K=1.
\end{equation*}
We have checked this property numerically for $K\le 4$ (see subsection \ref{factdimless}).
To prove it analytically from \eqref{this-is-the-end-beautiful-friend}
one has to carefully analyze the sub-leading corrections to the
integral equation \eqref{hdef} with the weight function \eqref{fpp}. We
leave this problem for further research. 

\subsection{The $c\to\infty$ limit}

Here we derive the leading term in the large coupling expansion of $g_K$.
In the $c\to\infty$ limit the kernel $\varphi(p)$ is of order $1/c$, therefore to
leading order
\begin{equation*}
  h^{(l)}(p)=p^l
\end{equation*}
and
\begin{equation*}
  \vev{\ordo_K}\quad\to\quad \frac{K!}{c^{K(K-1)}}
 \int \frac{dp_1}{2\pi}\dots \frac{d p_K}{2\pi}
\prod_{o} f(p_o)   \prod_{j>l}(p_j-p_l)^2.
\end{equation*}
Evaluating this formula for the ground state gives
\begin{equation*}
  g_K= \frac{K!}{2^K}\left(\frac{\pi}{\gamma}\right)^{K(K-1)}\times 
\int_{-1}^1 dx_1\dots dx_K \prod_{j>l} (x_j-x_l)^2 .
\end{equation*}
This result was already obtained in the papers \cite{GaShly-gK-2,GaShly-gK-1,sinhG-LL1,sinhG-LL2}.

\section{Factorization of the multiple integrals}

\label{factorization}

In this section we perform the factorization of the multiple integral
formula \eqref{ordoK} in the cases $K=2,3,4$ \footnote{The case $K=1$ is
trivial and it simply yields the particle density as it should.}. 
In order to keep the formulas as short as possible we will use the
following notation:
\begin{equation*}
  \int d\tilde p\ \cdots=\int \frac{dp}{2\pi} \ f(p)\ \cdots
\end{equation*}
Moreover we will suppress the dependence of the mean value on the
state $\ket{\Omega}$ and we write simply $\vev{\ordo_K}$. The
dependence on $\ket{\Omega}$ is carried by the functions $f(p)$ and $h^{(l)}(p)$.

We found it more convenient to work with the dimensionful formula
\eqref{ordoK}, because this way non-trivial checks of dimensional
analysis can be performed at each step of the calculation. The
dimensionless formulas can be obtained as explained in \ref{factdimless}.

The main idea behind the factorization procedure is simple: at each
step the number of the integrals can be reduced by one using the
integral equation \eqref{hdef}, whenever the prefactors are such that
the corresponding variable is present only in one denominator
\begin{equation*}
\frac{1}{(p_j-p_k)^2+c^2}= \frac{1}{2c} \varphi(p_j-p_k).
\end{equation*}
Except from the case $K=2$ this is not the case, instead
the prefactors have to be divided into several terms, in each of which
one of the integrals can be performed. This is a non-trivial task with
growing complexity as $K$ increases. In
the following we present a case-by-case study up until $K=4$.

\subsection{$K=2$}

One has
\begin{equation*}
   \vev{\ordo_2}=4\left[\int d\tilde p_1 d\tilde p_2 \frac{p_2}{(p_1-p_2)^2+c^2}
   h^{(0)}(p_1)h^{(1)}(p_2)-
\int d\tilde p_1 d\tilde p_2 \frac{p_1}{(p_1-p_2)^2+c^2}
   h^{(0)}(p_1)h^{(1)}(p_2)\right].
\end{equation*}
In the first term one integrates over $p_1$ first, in the second term
over $p_2$ first. Using the integral equation \eqref{hdef} one gets
\begin{equation}
\label{g2}
\begin{split}
  \vev{\ordo_2}&=\frac{2}{c}\left[
\int d\tilde p_2\ p_2 
   h^{(1)}(p_2) (h^{(0)}(p_2)-1)-
\int d\tilde p_1\ p_1 
   h^{(0)}(p_1) (h^{(1)}(p_1)-p_1)
\right]=\\
&=\frac{2}{c}\int d\tilde p\   ( p^2h^{(0)}(p)-ph^{(1)}(p)).
\end{split}
\end{equation}
This is in agreement with formula (10) of \cite{g3-exact}. As it was
already explained in \cite{sinhG-LL1,sinhG-LL2,g3-exact}, in the finite
temperature case \eqref{g2} agrees with the result obtained from the
Hellmann-Feynmann theorem. We also note that \eqref{g2} can be proven
for arbitrary weight function $f(p)$ using the Hellmann-Feynmann for a
single state; one has to repeat the arguments of Appendix D of
\cite{LM-sajat}. 

For future use we define
\begin{equation}
  \{n,m\}= \int d\tilde p\    p^n h^{(m)}(p)=
\int \frac{d p}{2\pi} f(p)\   p^n h^{(m)}(p).
\end{equation}
It can be shown using the iterative solution to \eqref{hdef} that in
general
\begin{equation*}
  \{n,m\}=\{m,n\}.
\end{equation*}
Using this notation we write
\begin{equation}
\label{J2}
  \vev{\ordo_2}=\frac{2}{c} \Big( \{0,2\}-\{1,1\}\Big)
\Big).
\end{equation}
Also, it is useful to derive a general formula which will be used often:
\begin{equation}
  \label{kettesint}
  \int d\tilde x d\tilde y \frac{x^\alpha-y^\alpha}{(x-y)^2+c^2}
  (h^{(\beta)}(x)h^{(\gamma)}(y)-h^{(\beta)}(y)h^{(\gamma)}(x))
  =\frac{1}{c}\Big( \{\gamma,\alpha+\beta\}-\{\beta,\alpha+\gamma\}\Big).
\end{equation}

\subsection{$K=3$} 

The mean value is given by
\begin{equation}
\label{k3m}
  \vev{\ordo_3}=36 \int {d\tilde p_1}\dots {d \tilde p_3}
 \prod_{j>l}\frac{p_j-p_l}{(p_j-p_l)^2+c^2}
 h^{(0)}(p_1)h^{(1)}(p_2)h^{(2)}(p_3).
\end{equation}
In this form neither of the integrals can be done directly. Instead,
one has to divide the prefactors into several terms such that in each of
them one integral can be performed. One way to do this is as
follows. We define
\begin{equation*}
  D(x,y,z)=\frac{y-x}{(x-y)^2+c^2}\frac{1}{(y-z)^2+c^2}.
\end{equation*}
Then we find the identity
\begin{equation}
\label{elso-azonossag}
\begin{split}
& \prod_{j>l}\frac{p_j-p_l}{(p_j-p_l)^2+c^2}=\\
&\frac{1}{3}\left[
D(x,y,z)+D(y,z,x)+D(z,x,y)-D(y,x,z)-D(x,z,y)-D(z,y,x)
\right] .
\end{split}
\end{equation}
For simplicity we used the variables $x,y,z$ on the r.h.s. instead of
$p_1,p_2,p_3$. Equation \eqref{elso-azonossag} can be proven as
follows. The r.h.s. is a completely anti-symmetric function of the
variables $x,y,z$ and it has exactly the same poles as the function on
the l.h.s., therefore it has to be equal to the
l.h.s. multiplied by a symmetric polynomial. By power counting it is
shown that this polynomial is a pure number. This
number is found to be $1$ by simple manipulations after sending $c\to 0$ on both sides.

Substituting \eqref{elso-azonossag} into \eqref{k3m}, performing one
integral in each term using 
\eqref{hdef}, changing variables accordingly 
and observing the cancellation of the terms including three $h$
functions we find
\begin{equation}
\label{g3koztes1}
\begin{split}
  \vev{\ordo_3}=\frac{6}{c} \int d\tilde x d\tilde y \frac{y-x}{(x-y)^2+c^2}\times
[ & y^2 (h_1(x)h_0(y)-h_0(x)h_1(y))+\\
 & y (h_0(x)h_2(y)-h_2(x)h_0(y))+\\
 &  (h_2(x)h_1(y)-h_1(x)h_2(y))
].
\end{split}
\end{equation}
Using \eqref{kettesint} the last line of \eqref{g3koztes1} yields
\begin{equation*}
  \frac{6}{c} \int d\tilde x d\tilde y \frac{y-x}{(x-y)^2+c^2}
  (h_2(x)h_1(y)-h_1(x)h_2(y))=
\frac{6}{c^2}\Big(\{2,2\}-\{1,3\}\Big).
\end{equation*}
For the second line of  \eqref{g3koztes1}  we can drop the term proportional to $xy$ to find
\begin{equation*}
\begin{split}
 \frac{6}{c} \int d\tilde x d\tilde y \frac{y^2}{(x-y)^2+c^2}
  (h_0(x)h_2(y)-h_2(x)h_0(y))=
\frac{3}{c^2}\Big(\{0,4\}-\{2,2\}\Big).
\end{split}
\end{equation*}
The first line of  \eqref{g3koztes1} is more involved. We have to compute
\begin{equation}
  \frac{3}{c} \int d\tilde x d\tilde y \frac{(y-x)(y^2+x^2)}{(x-y)^2+c^2}
  (h_1(x)h_0(y)-h_0(x)h_1(y)).
\end{equation}
We write
\begin{equation}
\label{koztes}
  \frac{(y-x)(y^2+x^2)}{(x-y)^2+c^2}=\frac{y-x}{3}+\frac{c^2}{3}\frac{x-y}{(x-y)^2+c^2}
+\frac{2}{3}\frac{y^3-x^3}{(x-y)^2+c^2}.
\end{equation}
The first term in \eqref{koztes} gives
\begin{equation*}
 \frac{2}{c}\Big(\{0,1\}^2-\{0,0\}\{1,1\}\Big).
\end{equation*}
The second and third terms in \eqref{koztes} can be evaluated using \eqref{kettesint}.

Putting everything together
\begin{equation}
  \label{J3}
\begin{split}
  \vev{\ordo_3}=&\frac{1}{c^2}  \Big(-4 \{1,3\}+3\{2,2\}+\{0,4\}\Big)
+\Big(\{0,2\}-\{1,1\}\Big)
+\frac{2}{c}\Big(\{0,1\}^2-\{0,0\}\{1,1\}\Big).
\end{split}
\end{equation}
This is in accordance with formula (11) of \cite{g3-exact}.

\subsection{$K=4$}

We define
\begin{equation*}
  D_4(x,y,z,u)=\frac{1}{((x-y)^2+c^2)((y-z)^2+c^2)((z-u)^2+c^2)}.
\end{equation*}
Then we find
\begin{equation}
\label{masodik-azonossag}
 \prod_{j>l}\frac{p_j-p_l}{(p_j-p_l)^2+c^2}=
\frac{1}{12}\sum_{\Pe \in \sigma_4} (-1)^{[\Pe]} D_4(\Pe p).
\end{equation}
This equation can be proven through the same steps as in the case of \eqref{elso-azonossag}.

One has to evaluate
\begin{equation*}
\begin{split}
  \vev{\ordo_4}&=24^2 \int {d\tilde p_1}\dots {d \tilde p_4}
 \prod_{j<l}\frac{p_j-p_l}{((p_j-p_l)^2+c^2)}
 h^{(0)}(p_1)h^{(1)}(p_2)h^{(2)}(p_3)h^{(3)}(p_4)=\\
&=48\int {d\tilde p_1}\dots {d \tilde p_4}
\frac{1}{((p_1-p_2)^2+c^2)((p_2-p_3)^2+c^2)((p_3-p_4)^2+c^2)} 
\sum_{\Pe \in \sigma_4} (-1)^{[\Pe]} \prod_{l=1}^4 h^{(\Pe_l-1)}(p_l).
\end{split}
\end{equation*}
Performing the integrals over $p_1$ and $p_4$ and using the symmetries
one gets
\begin{equation*}
  \vev{\ordo_4}=\frac{12}{c^2} \sum_{\Pe \in \sigma_4}(-1)^{[\Pe]} \int d\tilde p_2 d\tilde p_3
  \frac{p_2^{\Pe_1-1}p_3^{\Pe_4-1}}{(p_2-p_3)^2+c^2}
  h^{(\Pe_2-1)}(p_2) h^{(\Pe_3-1)}(p_3).
\end{equation*}
There are in total 6 different combinations and we treat them one by
one. We define
\begin{equation*}
 \vev{\ordo_4}=\frac{12}{c^2}\sum_{o=1}^6 K_o.
\end{equation*}
with $K_o$ representing one of the six combinations. 
The first three cases are evaluated easily:
\begin{equation*}
 K_1= \int d\tilde x d\tilde y \frac{y-x}{(x-y)^2+c^2} 
\big(h^{(2)}(x)h^{(3)}(y)-h^{(3)}(x)h^{(2)}(y)\big)
=\frac{1}{c}\Big(\{2,4\}-\{3,3\}   \Big)
\end{equation*}
\begin{equation*}
 K_2= \int d\tilde x d\tilde y \frac{y^2-x^2}{(x-y)^2+c^2} 
\big(h^{(3)}(x)h^{(1)}(y)-h^{(1)}(x)h^{(3)}(y)\big)
=\frac{1}{c}\Big(\{3,3\}-\{1,5\}   \Big)
\end{equation*}
\begin{equation*}
 K_3= \int d\tilde x d\tilde y \frac{y^3-x^3}{(x-y)^2+c^2} 
\big(h^{(1)}(x)h^{(2)}(y)-h^{(2)}(x)h^{(1)}(y)\big)
=\frac{1}{c}\Big(\{1,5\}-\{2,4\}   \Big)
\end{equation*}
The remaining three case are more complicated because we have to separate factors of the form
\begin{equation*}
  \frac{x^\alpha y^\beta-x^\beta y^\alpha}{(x-y)^2+c^2}\qquad\qquad \alpha,\beta>0.
\end{equation*}
The next case is
\begin{equation*}
K_4=  \int d\tilde x d\tilde y \frac{xy^2-yx^2}{(x-y)^2+c^2}
 \big(h^{(0)}(x)h^{(3)}(y)-h^{(3)}(x)h^{(0)}(y)\big).
\end{equation*}
Here we use
\begin{equation*}
    \frac{x y^2-x^2 y}{(x-y)^2+c^2}
=\frac{x-y}{3}+\frac{c^2}{3}\frac{y-x}{(x-y)^2+c^2}
+\frac{1}{3}\frac{y^3-x^3}{(x-y)^2+c^2}
\end{equation*}
leading to
\begin{equation*}
  K_4=\frac{2}{3}\Big(\{0,1\}\{0,3\}-\{0,0\}\{1,3\}\Big)+\frac{c}{3}
\Big(\{0,4\}-\{1,3\}\Big)+\frac{1}{3c} \Big(\{0,6\}-\{3,3\}\Big).
\end{equation*}
The next case is
\begin{equation*}
 K_5= \int d\tilde x d\tilde y \frac{xy^3-yx^3}{(x-y)^2+c^2}
 \big(h^{(2)}(x)h^{(0)}(y)-h^{(0)}(x)h^{(2)}(y)\big).
\end{equation*}
Here we use
\begin{equation*}
   \frac{x y^3-x^3 y}{(x-y)^2+c^2}=\frac{1}{2}\left[
x^2-y^2-c^2\frac{x^2-y^2}{(x-y)^2+c^2}+\frac{y^4-x^4}{(x-y)^2+c^2}
\right]
\end{equation*}
which gives
\begin{equation*}
  K_5=\Big(\{2,2\}\{0,0\}-\{0,2\}^2\Big)
+\frac{c}{2}\Big(\{2,2\}-\{0,4\}\Big)+
\frac{1}{2c}\Big(\{2,4\}-\{0,6\}\Big).
\end{equation*}
Finally, the last term is
\begin{equation*}
 K_6= \int d\tilde x d\tilde y \frac{x^2y^3-y^2x^3}{(x-y)^2+c^2} 
\big(h^{(0)}(x)h^{(1)}(y)-h^{(1)}(x)h^{(0)}(y)\big).
\end{equation*}
Here we write
\begin{equation*}
  \begin{split}
    \frac{x^2 y^3-x^3 y^2}{(x-y)^2+c^2}& =\frac{1}{5}
\left[\frac{2c^2}{3}(x-y)+(x^3-y^3)+2(x^2y-y^2x)
+\right.\\
&\left.+\frac{2c^4}{3}\frac{y-x}{(x-y)^2+c^2}
+\frac{5c^2}{3}\frac{y^3-x^3}{(x-y)^2+c^2}+\frac{y^5-x^5}{(x-y)^2+c^2}
\right]
  \end{split}
\end{equation*}
leading to
\begin{equation*}
\begin{split}
  K_6=&\frac{4c^2}{15}\Big(\{0,1\}^2-\{0,0\}\{1,1\}\Big)+
\frac{2}{5}\Big(\{0,3\}\{0,1\}-\{1,3\}\{0,0\}\Big)+\\
&\frac{4}{5}\Big(\{0,2\}\{1,1\}-\{0,1\}\{1,2\}\Big)+
\frac{2c^3}{15}\Big(\{0,2\}-\{1,1\}\Big)+
\frac{c}{3} \Big(\{0,4\}-\{1,3\}\Big)+\\
&\frac{1}{5c} \Big(\{0,6\}-\{1,5\}\Big).
\end{split}
\end{equation*}
Putting everything together
\begin{equation}
  \label{J4}
\begin{split}
\vev{\ordo_4}=\frac{2}{5c^3}\Big[&
8 c^3 \Big(\{0, 1\}^2  -  \{0, 0\} \{1, 1\}\Big) +  
32 c \Big(\{0, 1\} \{0, 3\}-  \{0, 0\} \{1, 3\}\Big) +\\
& 24c\Big(\{0, 2\} \{1, 1\}- \{0, 1\}\{1,2\}\Big) 
+ 30 c \Big(\{0, 0\} \{2, 2\}-\{0, 2\}^2\Big)+\\
& 4 c^4 \Big(\{0, 2\} -\{1,1\}\Big) + 
   5c^2 \Big(\{0, 4\}  - 4 \{1, 3\}+3  \{2, 2\}\Big)+\\
&  \{0, 6\}   - 6 \{1, 5\} + 15 \{2, 4\}  -  10 \{3, 3\} 
\Big].
\end{split}
\end{equation}
This a new result of the present work.

\subsection{Galilei  invariance}

The expectation values $\vev{\ordo_K}$ are Galilei invariant, and it
is useful to check this property in our final formulas. This
constitutes a highly non-trivial check, as it was already remarked in
\cite{g3-exact}. 

In our calculations we did not restrict ourselves to symmetric
distributions, the weight functions $f(p)$ can be
arbitrary. Therefore, to check Galilei invariance it is enough to
consider an infinitesimal boost $b$. This boost yields the following
infinitesimal transformations:
\begin{equation*}
  p\quad\to \quad p+b\qquad\qquad\qquad h^{(j)}(p)\quad\to\quad h^{(j)}(p)+b jh^{(j-1)}(p).
\end{equation*}
It is then readily seen that \eqref{ordoK} is invariant due to the
anti-symmetry of the prefactors.

We also performed the check on our factorized formulas. The
transformation rules for the quantities $\{\alpha,\beta\}$ are 
\begin{equation*}
  \{\alpha,\beta\}\quad\to\quad  \{\alpha,\beta\}+b\Big[
\alpha \{\alpha-1,\beta\}+\beta \{\alpha,\beta-1\}\Big].
\end{equation*}
Using this rule we have checked that the variation of equations
\eqref{J2}, \eqref{J3} and \eqref{J4} indeed vanishes.

\subsection{Dimensionless formulas and numerical results}

\label{factdimless}

The dimensionless versions of formulas \eqref{J2}, \eqref{J3} and
\eqref{J4} are obtained simply by setting $c=1$, multiplying with an
overall factor of $\gamma^K$, and replacing
\begin{equation}
  \{n,m\}\quad\to\quad \int d\tilde q\    q^n \tilde h^{(m)}(q),
\end{equation}

To numerically evaluate the factorized formulas the following steps
have to be performed:~
\begin{itemize}
\item Solve the TBA equation \eqref{dimlessTBA} iteratively. The
  parameter $\alpha$ can be fixed by requiring
  \begin{equation*}
    g_1=\gamma \{0,0\}=1
  \end{equation*}
\item Solve the linear integral equations \eqref{qdef} for $\tilde
  h^{(l)}(q)$.
\item Evaluate \eqref{J2}, \eqref{J3} and \eqref{J4}.
\end{itemize}
We peformed this procedure for 
a wide range of the parameters $\gamma$ and $\tau$. The quantity $g_4$ 
shows the same qualitative behaviour as $g_2$ and $g_3$ 
\cite{g3-exact}: it is an increasing function of $\tau$ and a
decreasing function of $\gamma$, with the limiting values given by
\begin{equation*}
  \lim_{\gamma\to 0} g_4=\lim_{\tau\to\infty} g_4=4!=24\qquad
  \qquad\lim_{\gamma\to 0}\lim_{\tau\to 0} g_4=1.
\end{equation*}

\begin{figure}
  \centering
\psfrag{g2}{$g_2$}
\psfrag{g3}{$g_3$}
\psfrag{g4}{$g_4$}
\psfrag{logg}{$\log_{10}\gamma$}
  \includegraphics[scale=1]{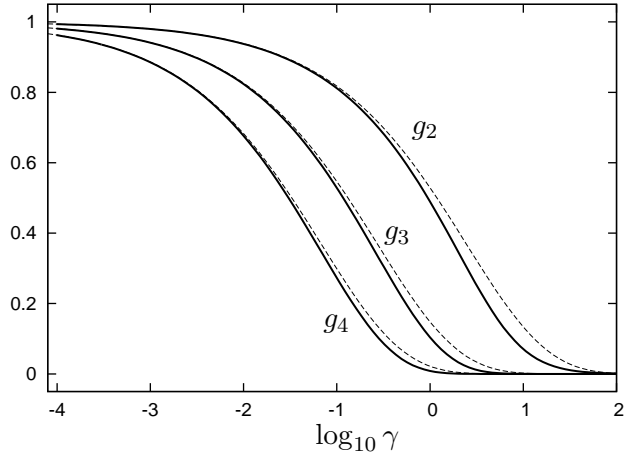}
  \caption{The ground state values of the $K$-body local correlations
    for $K\le 4$ as a function of the
    dimensionless coupling $\gamma$  ($g_1=1$ by definition). The
    exact values are represented by the solid lines, whereas the
    dashed lines show the empirical formula \eqref{empirikus}.}
  \label{gK-T0}
\end{figure}

\begin{figure}
  \centering
\psfrag{tau01}{$\gamma=0.1$}
\psfrag{tau1}{$\gamma=1$}
\psfrag{tau10}{$\gamma=10$}
\psfrag{log10tau}{$\log_{10}\tau$}
  \includegraphics[scale=1]{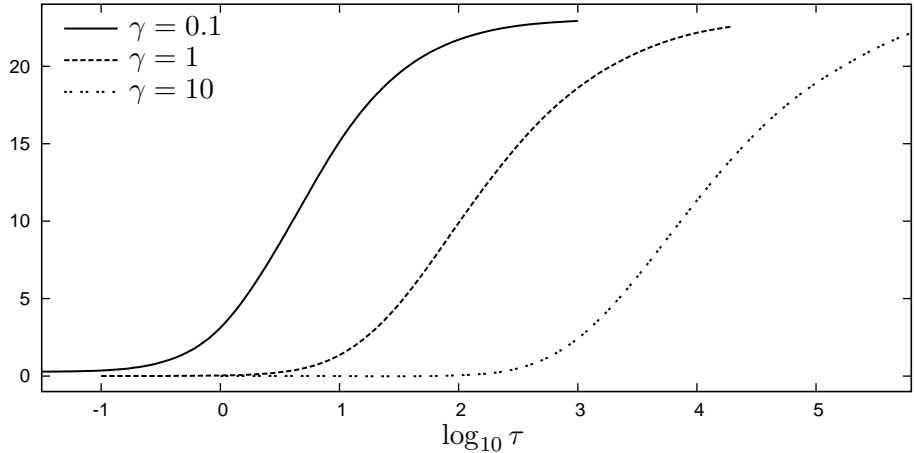}
  \caption{The quantity $g_4$ as a function of the dimensionless
    temperature $\tau$ for intermediate couplings. In the $\tau\to 0$
    limit the three curves approach small (but non-zero) values which
    are shown in Figure \ref{gK-T0}.  }
  \label{gK-gamma}
\end{figure}

To demonstrate the numerical
results we present the ground state values of $g_2$, $g_3$ and $g_4$
in Fig. \ref{gK-T0}, whereas the temperature dependence of $g_4$ is
shown in Fig. \ref{gK-gamma} for the intermediate couplings
$\gamma=0.1$, $\gamma=1$ and $\gamma=10$. 

At $T=0$ the first term in the small coupling expansion of $g_K$ is
given by \cite{GaShly-gK-1}:
\begin{equation*}
g_K=1-\frac{K(K-1)}{\pi}\sqrt{\gamma}+\ordo(\gamma)  
\end{equation*}
We found that the empirical formula 
\begin{equation}
\label{empirikus}
 g_K\approx \exp\left(-\frac{K(K-1)}{\pi}\sqrt{\gamma}\right)
\end{equation}
gives a surprisingly good approximation and can be used for practical purposes
even at $\gamma\sim 1$. The predictions of \eqref{empirikus} are also
plotted in Fig. \ref{gK-T0}. It is expected that \eqref{empirikus}
holds with a good approximation even for higher $K$.

It would be useful to compare the exact numerical values for $g_3$ and
$g_4$ case to the various approximations available
in the literature \cite{GaShly-gK-2,GaShly-gK-1} including the
large-coupling expansion both at zero and finite temperatures. 
This is out of the scope of the present work and is left for further
research.

\newpage

\section{Mean values in the LeClair-Mussardo formalism}

In this section we elaborate on 
the LeClair-Mussardo formalism, which is an alternative approach to
obtain expectation values of local operators leading to an infinite
integral series
\cite{leclair_mussardo,fftcsa2,sinhG-LL1,sinhG-LL2,LM-sajat}. 
For our present purposes the following form of the series is the most
convenient \cite{LM-sajat}:
\begin{equation}
  \begin{split}
 \vev{\ordo_K}=
\sum_N \frac{1}{N!}
\int \frac{dp_1}{2\pi}\dots  \frac{dp_N}{2\pi}
\left( \prod_{j=1}^N f(p_j)\omega(p_j)\right)
F^K_{N,s}(p_1,\dots,p_N),
\label{LMsajat}
  \end{split}
\end{equation}
where
\begin{equation*}
  \omega(p)=\exp\left(-\int \frac{dp'}{2\pi} f(p') \varphi(p-p')\right)
\end{equation*}
and $f(p)$ is defined in \eqref{fp}. The form factors appearing in the
above series are defined as
\begin{equation}
\label{FKNS}
  F^K_{N,s}(p_1,\dots,p_N)=\prod_{j<k}\frac{(p_j-p_k)^2}{(p_j-p_k)^2+c^2}\times
\lim_{\eps\to 0} F^K_N({\{p_j+\eps\}},\{p_j\}),
\end{equation}
where the form factor on the r.h.s. is given by \eqref{FFK}. This
prescription is also called the ``symmetric evaluation of the diagonal limit'';
note that this limit is different from the way we
obtained the mean value \eqref{hat-ez-mar-a-sokadik-tipp} because in \eqref{FFK} the
Bethe equations were substituted into the matrix element {\it before}
taking the diagonal limit. Therefore the object in \eqref{FKNS} does not depend
on the volume $L$. Note also that the l.h.s. refers to a normalization where the norm of
the Bethe state is given simply by the Gaudin-determinant
$\mathcal{G}^{LL}$. 

Alternatively \eqref{LMsajat} can be expressed as \cite{LM-sajat,sinhG-LL1,sinhG-LL2}
\begin{equation}
\label{LM}
 \vev{\ordo_K}=
\sum_N \frac{1}{N!}
\int \frac{dp_1}{2\pi}\dots  \frac{dp_N}{2\pi}
\left( \prod_{j=1}^N f(p_j)\right)
F^K_{N,c}(p_1,\dots,p_N).
\end{equation}
Here $F^K_{N,c}(p_1,\dots,p_N)$ are the so-called connected
evaluations of the diagonal form factors. Their precise definition and
the relation to $F^K_{N,s}(p_1,\dots,p_N)$ can be found in
\cite{fftcsa2,LM-sajat}. The series \eqref{LM}  was originally
developed in \cite{leclair_mussardo} in the framework of integrable
Quantum Field Theories. Later it was used in \cite{sinhG-LL1,sinhG-LL2,g3-exact} to
compute the quantities $g_K$ up to $K~=~3$. 
However, for higher $K$
 the results \eqref{LMsajat}-\eqref{LM} are only 
formal because the form factors themselves were not calculated previously.
{In \cite{sinhG-LL1,sinhG-LL2} a prescription was given of
  how to obtain the connected evaluation using a special non-relativistic
  limit of certain form factors of the sinh-Gordon model. However,
  the actual calculation becomes more and more demanding
  with higher $K$ and $N$.}

We
fill this gap here by calculating the explicit results for
$F^K_{N,s}(p_1,\dots,p_N)$ for arbitrary $K$ and $N$: we take the
symmetric diagonal limit of the form factor 
\eqref{FFK}. 
Note that every singularity of the form
factor is
included in the matrix $Z$, and even the elements of $Z$ can be
evaluating easily. Taking the limit and multiplying with the
prefactors we obtain
\begin{equation}
\label{elso-tipp}
  F^K_{N,s}(p_1,\dots,p_m)=(K!)^2
\mathop{\sum_{\{p^+\}\cup \{p^-\}}}_{|\{p^+\}|=K} 
\left[\prod_{j>l}\frac{p_j^+-p_l^+}{(p_j^+-p_l^+)^2+c^2)}\right]
\times \det Y.
\end{equation}
The elements of $Y$ are given by 
\begin{equation}
\label{dub-trees}
  \begin{split}
 Y_{jl}&=(p_j)^{l-1}
 \qquad\text{if}\quad p_l\in \{p^+\}\\
  Y_{j,l}&=\delta_{j,l}\Big(\sum_{o=1}^N \varphi(p_j-p_o) \Big)-\varphi(p_j-p_l)
\qquad\text{if}\quad p_l\in \{p^-\}.
  \end{split}
\end{equation}
Note that
\begin{equation*}
  Y=\mathcal{H}\Big|_{L=0},
\end{equation*}
where $\mathcal{H}$ is the matrix defined in \eqref{maia}.

With these results the series \eqref{LMsajat} can be considered an
explicit representation of the mean value. 

It would be desirable to
have a general recipe for the re-summation of the series, which would
be an alternative way to obtain factorized formulas like \eqref{J3} and \eqref{J4}.
However, this is far from being easy. 
The simpler cases $K=1$ and $K=2$ were already calculated in
\cite{sinhG-LL1,sinhG-LL2}. The highly non-trivial case of $K=3$ was
considered in \cite{g3-exact}, where the authors evaluated the series
\eqref{LM} (and obtained the result \eqref{J3} for the first time)
based on the following conjecture for the quantities $F^3_{N,c}$ 
\footnote{In
  order to ensure compatibility with our normalizations we inserted a
  factor of $1/c^2$.}:
\begin{equation}
\label{marcieke3}
  F^{3}_{N,c}=\frac{1}{2c^2}\sum_P \varphi_{12}\varphi_{23}\dots \varphi_{N-1,N} p_{1N}
(p_{1N}^3-p_{12}^3-p_{23}^3-\dots-p_{N-1,N}^3).
\end{equation}
Here we check this formula in the first two cases. 
In the simplest case of $N=3$ our formula \eqref{elso-tipp} gives
\begin{equation*}
  F^3_{3,s}=F^3_{3,c}=36 \prod_{j>l} \frac{p_{jl}^2}{p_{jl}^2+c^2}.
\end{equation*}
This was already calculated in \cite{sinhG-LL1,sinhG-LL2} and is in agreement with
\eqref{marcieke3}. In the case of $N=4$
 one has to use
the following relation between the symmetric and connected evaluations
\cite{fftcsa2}:
\begin{equation*}
  F^3_{4,s}(p_1,p_2,p_3,p_4)=F^3_{4,c}(p_1,p_2,p_3,p_4)+\sum_j
  F^3_{3,c}(\hat p_j)\times \left(\sum_{k\ne j} \varphi_{jk}\right).
\end{equation*}
Here $\hat p_j$ means that $p_j$ is not present among the arguments of
the form factor. We used the program \texttt{Mathematica} to express $F^3_{4,c}$ using
the above relation and we found agreement with \eqref{marcieke3}. This
is a highly non-trivial check of the conjecture \eqref{marcieke3}; a
 proof for arbitrary $N$ is not known.

Finally we note that in the simpler cases of 
$K=1$ and $K=2$ we evaluated \eqref{elso-tipp} and found exact agreement with the corresponding
formulas of Appendix D in  \cite{LM-sajat}.

\section{Conclusions}

We developed multiple integral formulas for the local correlations in
the 1D Bose gas. The final results for the expectation value
$\vev{\ordo_K}$ is given by equation \eqref{ordoK}, whereas the
dimensionless formula for $g_K$ is given by \eqref{gK}.

In section \ref{factorization} we performed the explicit factorization
of the multiple integrals in the cases $K=2,3,4$; for $K=3$ we obtained
the recent result of \cite{g3-exact} whereas our formula for $K=4$
is new. Our method of factorization relies only on the integral
equation \eqref{hdef} defining the auxiliary functions entering the
multiple integral. 
Therefore the process works
for arbitrary distribution of Bethe roots and not only for the ground
state or the finite temperature Gibbs states.

The general recipe of how to perform the
factorization for $K>4$ is not known. The strategy is clear:
at each step the prefactors have to be manipulated in such a way that
the number of integrals can be reduced by one using the integral
equation \eqref{hdef}. We believe that this can always be done and it
would be interesting to develop a general algorithm for this process.

An alternative way to obtain the mean values $\vev{\ordo_K}$ would be
to take the thermodynamic limit on the XXZ spin chain first and to perform
the scaling limit towards the Bose gas afterwards. The advantage of
this approach would be that on the spin chain the factorization of the
multiple integral formulas for the elements of the reduced density
matrix is by now well-understood (see
 \cite{XXZ-factorization-recent-osszefoglalo} and references
therein). In fact we attempted to take the scaling limit of the
factorized results of \cite{XXZ-finite-T-factorization} concerning
the emptiness formation probability. However, this turned out to be 
cumbersome already in the case $K=2$. Thus it seems that the direct
approach of the present paper is more advantageous, at least for the
small values of $K$ considered here.

An other alternative way would be to sum up the LeClair-Mussardo
series \eqref{LMsajat} or \eqref{LM}. The diagonal form factors
entering \eqref{LMsajat} are given explicitly by \eqref{elso-tipp},
therefore the remaining task is purely combinatorial: one has to expand
the sums of determinants appearing in \eqref{elso-tipp} and put the
resulting expression in a form which is amenable for
re-summation. Again, this is a formidable problem, the solution of
which is not yet known.

\vspace{1cm}
{\bf Acknowledgements} The author is grateful to M. Kormos and
A. Imambekov for inspiring discussions and for communicating the
results of \cite{g3-exact} prior to publication. Also, we are indebted
to J. Mossel, N. J. van Druten, J-S. Caux and M. Kormos for useful
comments about the manuscript.

\addcontentsline{toc}{section}{References}
\bibliography{XXZ-to-LL}
\bibliographystyle{utphys}

\end{document}